\documentclass[preprintnumbers,amsmath,amssymbm,prd]{revtex4}
\usepackage{epsfig}
\usepackage{graphicx}
\usepackage{amssymb}

\begin{document}
\title{Eigenvalue spectrum of the spheroidal harmonics: a uniform asymptotic analysis}
\author{Shahar Hod}
\address{The Ruppin Academic Center, Emeq Hefer 40250, Israel}
\address{ }
\address{The Hadassah Institute, Jerusalem 91010, Israel}
\date{\today}

\begin{abstract}
\ \ \ The spheroidal harmonics $S_{lm}(\theta;c)$ have attracted the
attention of both physicists and mathematicians over the years.
These special functions play a central role in the mathematical
description of diverse physical phenomena, including black-hole
perturbation theory and wave scattering by nonspherical objects. The
asymptotic eigenvalues $\{A_{lm}(c)\}$ of these functions have been
determined by many authors. However, it should be emphasized that
all previous asymptotic analyzes were restricted either to the
regime $m\to\infty$ with a {\it fixed} value of $c$, or to the
complementary regime $|c|\to\infty$ with a {\it fixed} value of $m$.
A fuller understanding of the asymptotic behavior of the eigenvalue
spectrum requires an analysis which is asymptotically uniform in
{\it both} $m$ and $c$. In this paper we analyze the asymptotic
eigenvalue spectrum of these important functions in the {\it double}
limit $m\to\infty$ and $|c|\to\infty$ with a fixed $m/c$ ratio.
\end{abstract}
\bigskip
\maketitle

%]

\section{Introduction.}

The spheroidal harmonic functions $S(\theta;c)$ appear in many
branches of physics. These special functions are solutions of the
angular differential equation \cite{Flam,Teu,Abram}
\begin{equation}\label{Eq1}
{1\over {\sin\theta}}{\partial \over
{\partial\theta}}\Big(\sin\theta {{\partial
S}\over{\partial\theta}}\Big)+\Big[c^2\cos^2\theta-{{m^2}\over{\sin^2\theta}}+A\Big]S=0\
,
\end{equation}
where $\theta\in [0,\pi]$, $c\in\mathbb{Z}$, and the integer
parameter $m$ is the azimuthal quantum number of the wave field
\cite{Flam,Teu,Abram}.

These angular functions play a key role in the mathematical
description of many physical phenomena, such as: perturbation theory
of rotating Kerr black holes \cite{Teu,Ber1,Hod1,Notebh},
electromagnetic wave scattering \cite{Asa}, quantum-mechanical
description of molecules \cite{Eyr,Fal}, communication theory
\cite{Iee}, and nuclear physics \cite{BDB}.

The characteristic angular equation (\ref{Eq1}) for the spheroidal
harmonic functions is supplemented by a regularity requirement for
the corresponding eigenfunctions $S(\theta;c)$ at the two boundaries
$\theta=0$ and $\theta=\pi$. These boundary conditions single out a
{\it discrete} set of eigenvalues $\{A_{lm}\}$ which are labeled by
the discrete spheroidal harmonic index $l$ (where
$l-|m|=\{0,1,2,...\})$. For the special case $c=0$ the spheroidal
harmonic functions $S(\theta;c)$ reduce to the spherical harmonic
functions $Y(\theta)$, which are characterized by the familiar
eigenvalue spectrum $A_{lm}=l(l+1)$.

The various asymptotic spectrums of the spheroidal harmonics with
$c^2\in\mathbb{R}$ (when $c\in\mathbb{R}$ the corresponding
eigenfunctions are called oblate, while for $ic\in\mathbb{R}$ the
eigenfunctions are called prolate) were explored by many authors,
see \cite{BCC,Yang,Flam,Meix,Breu,Cas,Hodnw} and references therein.
In particular, in the asymptotic regime $m^2\gg |c|^2$ the
eigenvalue spectrum is given by \cite{BCC,Yang}
\begin{equation}\label{Eq2}
A_{lm}=l(l+1)-{{c^2}\over{2}}\Big[1-{{m^2}\over{l(l+1)}}\Big]+O(1)\
,
\end{equation}
while in the opposite limit, $|c|^2\gg m^2$ with $ic\in\mathbb{R}$,
the asymptotic spectrum is given by \cite{BCC,Flam,Meix,Breu,Hodnw}
\begin{equation}\label{Eq3}
A_{lm}=[2(l-m)+1]|c|+O(1)\  .
\end{equation}
The asymptotic regime $c^2\gg m^2$ (with $c\in\mathbb{R}$) was
studied in \cite{BCC,Flam,Meix,Breu,Cas,Hodnw,Notepm}, where it was
found that the eigenvalues are given by:
\begin{equation}\label{Eq4}
A_{lm}=-c^2+2[l+1-\text{mod}(l-m,2)]c+O(1)\  .
\end{equation}
Note that the spectrum (\ref{Eq4}) is doubly degenerate.

It should be emphasized that all previous asymptotic analyzes of the
eigenvalue spectrum were restricted either to the regime
$m\to\infty$ with a {\it fixed} value of $c$ \cite{BCC,Yang}, or to
the complementary regime $|c|\to\infty$ with a {\it fixed} value of
$m$ \cite{BCC,Flam,Meix,Breu,Cas}. A complete understanding of the
asymptotic eigenvalue spectrum requires an analysis which is uniform
in both $m$ and $c$ [that is, a uniform asymptotic analysis which is
valid for a fixed (non-negligible) $m/c$ ratio as {\it both} $m$ and
$|c|$ tend to infinity].

The main goal of the present paper is to present a uniform
asymptotic analysis for the spheroidal harmonic eigenvalues in the
{\it double} asymptotic limit
\begin{equation}\label{Eq5}
m\to\infty\ \ \ \text{and}\ \ \ |c|\to\infty
\end{equation}
with a fixed $m/c$ ratio.

\section{A transformation into the Schr\"odinger-type wave equation}
%{A coordinate transformation.}

For the analysis of the asymptotic eigenvalue spectrum, it is
convenient to use the coordinate $x$ defined by \cite{Yang,Hodnw}
\begin{equation}\label{Eq6}
x\equiv\ln\Big(\tan\Big({{\theta}\over{2}}\Big)\Big)\  ,
\end{equation}
in terms of which the angular equation (\ref{Eq1}) for the
spheroidal harmonic eigenfunctions takes the form of a
one-dimensional Schr\"odinger-like wave equation \cite{NoteSch}
\begin{equation}\label{Eq7}
{{d^2S}\over{dx^2}}-US=0\  ,
\end{equation}
where the effective radial potential is given by
\begin{equation}\label{Eq8}
U(x(\theta))=m^2-\sin^2\theta(c^2\cos^2\theta+A)\ .
\end{equation}
Note that the transformation (\ref{Eq6}) maps the interval
$\theta\in [0,\pi]$ into $x\in[-\infty,\infty]$.

The effective potential $U(\theta)$ is invariant under the
transformation $\theta\to\pi-\theta$. It is characterized by two
qualitatively different spatial behaviors depending on the relative
magnitudes of $A$ and $c^2$. We shall now study the asymptotic
behaviors of the spheroidal eigenvalues in the two distinct cases:
$A/c^2>1$ and $A/c^2<1$ \cite{Notebl}.

\section{The asymptotic eigenvalue spectrum}

\subsection{The asymptotic regime $\{|c|,m\}\to\infty\ $ with$\ \ c^2<m^2$.}

If $A>c^2$ then the effective radial potential $U(x(\theta))$ is in
the form of a symmetric potential well whose local minimum is
located at
\begin{equation}\label{Eq9}
{\theta_{\text{min}}}={\pi\over 2}\ \ \ \text{with} \ \ \
U({\theta_{\text{min}}})=-A+m^2\  .
\end{equation}
[Note that ${\theta_{\text{min}}}={\pi\over 2}$ corresponds to
$x_{\text{min}}=0$.]

Spatial regions in which $U(x)<0$ (the `classically allowed
regions') are characterized by an oscillatory behavior of the
corresponding wave function $S$, whereas spatial regions in which
$U(x)>0$ are characterized by an exponentially decaying wave
function (these are the `classically forbidden regions'). The
effective radial potential $U(x)$ is characterized by two `classical
turning points' $\{x^-,x^+\}$ (or equivalently,
$\{{\theta^-},{\theta^+}\}$) for which $U(x)=0$ \cite{Notetur}.

The one-dimensional Schr\"odinger-like wave equation (\ref{Eq7}) is
in a form that is amenable to a standard WKB analysis. In
particular, a standard textbook second-order WKB approximation
yields the well-known quantization condition
\cite{WKB1,WKB2,WKB3,Iyer,Notehigh}
\begin{equation}\label{Eq10}
\int_{x^{-}}^{x^{+}}dx\sqrt{-U(x)}=(N+{1\over 2})\pi\ \ \ ; \ \ \
N=\{0,1,2,...\}\
\end{equation}
for the bound-state `energies' (eigenvalues) of the
Schr\"odinger-like wave equation (\ref{Eq7}), where $N$ is a
non-negative integer. The characteristic WKB quantization condition
(\ref{Eq10}) determines the eigenvalues $\{A\}$ of the spheroidal
harmonic functions in the double limit $\{|c|,m\}\to\infty$. The
relation so obtained between the angular eigenvalues and the
parameters $m,c,$ and $N$ is rather complex and involves elliptic
integrals. However, if we restrict ourselves to the fundamental
(low-lying) modes which have support in a small interval around the
potential minimum ${x_{\text{min}}}$ \cite{Notereg1}, then we can
use the expansion $U(x)\simeq U_{\text{min}}+{1\over
2}U^{''}_{\text{min}}(x-x_{\text{min}})^2+O[(x-x_{\text{min}})^4]$
in (\ref{Eq10}) to obtain the WKB quantization condition \cite{Iyer}
\begin{equation}\label{Eq11}
{{|U_{\text{min}}|}\over{\sqrt{2U^{''}_{\text{min}}}}}=N+{1\over 2}\
\ \ ; \ \ \ N=\{0,1,2,...\}\ ,
\end{equation}
where a prime denotes differentiation with respect to $x$. The
subscript ``$\text{min}$" means that the quantity is evaluated at
the minimum $x_{\text{min}}$ of $U(x(\theta))$. Substituting
(\ref{Eq8}) with $x_{\text{min}}=0$ into the WKB quantization
condition (\ref{Eq11}), one finds the asymptotic eigenvalue spectrum
\begin{equation}\label{Eq12}
A(c,m,N)=m^2+(2N+1)\sqrt{m^2-c^2}+O(1)\ \ \ ; \ \ \ N=\{0,1,2,...\}\
\end{equation}
in the $N\ll \sqrt{m^2-c^2}$ regime \cite{Notereg1}. The resonance
parameter $N=\{0,1,2,...\}$ corresponds to $l-|m|=\{0,1,2,...\}$,
where $l$ is known as the spheroidal harmonic index.

It is worth noting that the eigenvalue spectrum (\ref{Eq12}), which
was derived in the {\it double} asymptotic limit
$\{|c|,m\}\to\infty$,
%with a fixed $c/m$ ratio
reduces to (\ref{Eq2}) in the special case $m\gg |c|$ and reduces to
(\ref{Eq3}) in the opposite special case $|c|\gg m$ with
$ic\in\mathbb{R}$. The fact that our uniform eigenvalue spectrum
(\ref{Eq12}) reduces to (\ref{Eq2}) and (\ref{Eq3}) in the
appropriate special limits provides a consistency check for our
analysis \cite{Notenum1}.

\subsection{The asymptotic regime $\{c,m\}\to\infty\ $ with$\ \ c^2>m^2$.}

If $A<c^2$ then the effective radial potential $U(x(\theta))$ is in
the form of a symmetric double-well potential: it has a local
maximum at
\begin{equation}\label{Eq13}
{\theta_{\text{max}}}={\pi\over 2}\ \ \ \text{with} \ \ \
U({\theta_{\text{max}}})=-A+m^2\  ,
\end{equation}
and two local minima at \cite{Notens}
\begin{equation}\label{Eq14}
\theta^{\pm}_{\text{min}}={1\over 2}\arccos(-A/c^2)
\end{equation}
with
\begin{equation}\label{Eq15}
%U(\theta^{\pm}_{\text{min}})=-{1\over
%4}c^2\Big[1-\Big({{A}\over{c^2}}\Big)^2\Big]-{1\over
%2}A\Big[1+{{A}\over{c^2}}\Big]+m^2\  .
U(\theta^{\pm}_{\text{min}})=-{1\over
4}c^2\big[1-({{A}/{c^2}})^2\big]-{1\over
2}A\big[1+({{A}/{c^2}})\big]+m^2\  .
\end{equation}
Thus, the two potential wells are separated by a large
potential-barrier of height
\begin{equation}\label{Eq16}
\Delta U\equiv
U(\theta_{\text{max}})-U(\theta^{\pm}_{\text{min}})={1\over
4}c^2\big[1-({{A}/{c^2}})^2\big]-{1\over
2}A\big[1-({{A}/{c^2}})\big] \to\infty\ \ \ \text{as}\ \ \
c\to\infty\  .
\end{equation}
The fact that the two potential wells are separated by an infinite
potential-barrier in the $c\to\infty$ limit (with $c^2>m^2$)
\cite{Notebar} implies that the coupling between the wells (the
`quantum tunneling' through the potential barrier) is negligible in
the $c\to\infty$ limit. The two potential wells can therefore be
treated as independent of each other in the $c\to\infty$ limit
\cite{WKB1,Zhou}. Thus, the two spectra of eigenvalues (which
correspond to the two identical potential wells) are degenerate in
the $c\to\infty$ limit \cite{Notetun}.

Substituting (\ref{Eq8}) with $\theta_{\text{min}}={1\over
2}\arccos(-A/c^2)$ into the WKB quantization condition (\ref{Eq11}),
one finds the asymptotic eigenvalue spectrum
\begin{equation}\label{Eq17}
A(c,m,N)=-c^2+2\big[m+(2N+1)\sqrt{1-{m/c}}\big]c+O(1)\ \ \ ; \ \ \
N=\{0,1,2,...\}\
\end{equation}
in the $N\ll m\sqrt{1-{m/c}}$ regime \cite{Notereg2}. We recall that
the spectrum (\ref{Eq17}) is doubly degenerate in the $c\to\infty$
regime \cite{Notedeg}; each value of $N$ corresponds to two adjacent
values of the spheroidal harmonic index $l$: $N={1\over
2}[l-m-\text{mod}(l-m,2)]$ \cite{Notenlm}.

It is worth noting that the eigenvalue spectrum (\ref{Eq17}), which
was derived in the {\it double} asymptotic limit
$\{|c|,m\}\to\infty$,
%with a fixed $c/m$ ratio
reduces to (\ref{Eq4}) in the special case $c^2\gg m^2$. The fact
that our uniform eigenvalue spectrum (\ref{Eq17}) reduces to
(\ref{Eq4}) in the appropriate special limit provides a consistency
check for our analysis \cite{Notenum2}.

\bigskip
\noindent
{\bf ACKNOWLEDGMENTS}
\bigskip

This research is supported by the Carmel Science Foundation. I thank
Yael Oren, Arbel M. Ongo, Ayelet B. Lata, and Alona B. Tea for
stimulating discussions.


\begin{thebibliography}{99}
\bibitem{Notebh} In the context of black-hole perturbation theory,
the dimensionless parameter $c$ stands for $a\omega$, where $a$ is
the specific angular momentum (angular momentum per unit mass) of
the spinning black hole and $\omega$ is the conserved frequency of a
scalar perturbation mode \cite{Teu}.
\bibitem{Notepm} We shall assume without loss of generality that $\Re c\geq0$, $\Im c\geq0$, and $m\geq0$.
Note that the angular differential
equation (\ref{Eq1}) is invariant under the transformations $c\to
-c$ and $m\to -m$. Thus, the eigenvalues are also invariant under
these transformations.
\bibitem{NoteSch} Note that in the quantum-mechanical terminology
$-U$ stands for ${{2m}\over{\hbar^2}}(E-V)$, where $E, V$, and $m$
are the total energy, potential energy, and mass of the of the
particle, respectively.
\bibitem{Notebl} Below we shall show
that these two cases correspond to $c^2<m^2$ and $c^2>m^2$,
respectively.
\bibitem{Notetur} Note that these turning points are characterized by the relation
${\theta^-}<{\theta_{\text{min}}}<{\theta^+}$.
\bibitem{Notehigh} Higher order corrections to the asymptotic eigenvalues [see formulas (\ref{Eq12})
and (\ref{Eq17}) below] can be obtained by using a higher-order WKB
analysis \cite{Iyer}.
\bibitem{Notereg1} Substituting our final formula [see Eq.
(\ref{Eq12}) below] into the effective potential (\ref{Eq8}), one
finds that the turning points are located at
${x^{\pm}}-{x_{\text{min}}}\simeq \pm \sqrt{{{A-m^2}\over{A-c^2}}}
%+O[(m^2-c^2)^{-3/2}]
\simeq\pm\sqrt{{{2N+1}\over{\sqrt{m^2-c^2}}}}$.
Thus, the assumption $|x^{\pm}-x_{\text{min}}|\ll 1$ is valid in the
$N\ll \sqrt{m^2-c^2}$ regime.
\bibitem{Notenum1} It is worth noting that our analytical formula
(\ref{Eq12}) agrees with the numerical results of \cite{Fal} for the
case $l=m=100$ with $c=100i$ with a remarkable accuracy of
$3.68\times 10^{-3}\%$. (Note that $c\to ic$ in the notations of
\cite{Fal}).
\bibitem{Notebar} Substituting our final formula [see Eq.
(\ref{Eq17}) below] into the effective potential (\ref{Eq8}), one
finds that the potential barrier (\ref{Eq16}) is given by $\Delta
U=(c-m)^2+O(c)$. Thus, $\Delta U\to\infty$ in the $c\to\infty$ limit
with $c^2>m^2$.
\bibitem{Notetun} More precisely, the coupling between the two potential wells (due to the
weak `quantum tunneling' through the large potential barrier)
introduces a small correction term of order
$\exp[{-\int_{\theta^-_2}^{\theta^+_1}d\theta\sqrt{U(\theta)}}]$ to
the r.h.s of the WKB quantization condition (\ref{Eq10})
\cite{Zhou,WKB1}, where ${\theta^-_2}$ and ${\theta^+_1}$ are the
inner turning points of the effective potential barrier. This term
is of the order of $e^{-\sqrt{\Delta U}}\sim e^{-(c-m)}\to 0$ [see
Eqs. (\ref{Eq16}) and (\ref{Eq17})] and is therefore negligible in
the $c\to\infty$ limit with $c>m$ \cite{Zhou,WKB1}.
\bibitem{Notereg2} Substituting our final formula (\ref{Eq17}) into the effective potential (\ref{Eq8}), one finds
that the turning points are located at
${x^{\pm}}-{x_{\text{min}}}\simeq \pm \sqrt{{{N+{1\over
2}}\over{m\sqrt{1-{m\over c}}}}}$. Thus, the assumption
$|x^{\pm}-x_{\text{min}}|\ll 1$ is valid in the $N\ll
m\sqrt{1-{m\over c}}$ regime.
\bibitem{Notedeg} As discussed above, this double degeneracy of the asymptotic eigenvalue
spectrum reflects the fact that the effective potential (\ref{Eq8})
with $c^2>m^2$ is composed of two identical potential wells which,
in the $c\to\infty$ limit, are separated by an infinite
potential-barrier.
\bibitem{Notenum2} It is worth noting that our analytical formula
(\ref{Eq17}) agrees with the numerical results of \cite{Fal} for the
case $l=m=100$ with $c=100$ with a remarkable accuracy of $0.22\%$.
(Note that $c\to ic$ in the notations of \cite{Fal}).
\end{thebibliography}
\end{document}